\documentclass[aps,prx,twocolumn,superscriptaddress,showkeys,english,10pt]{revtex4-2}
\usepackage[english]{babel}
\usepackage[utf8]{inputenc}
\usepackage{amsmath}
\usepackage{booktabs}
\usepackage{graphicx}
\usepackage{xcolor}
\usepackage{multirow}
\usepackage[normalem]{ulem} 
\usepackage{hyperref} 
\hypersetup{
colorlinks=true,
linkcolor=black,
citecolor=black,
filecolor=black,
urlcolor=black
}
\newcommand{\Angstrom}{\text{\AA}} 
\newcommand{\degree}{^{\circ}} 
\newcommand{\lowTpopulation}{ \xi } 

\newcommand{\Sconf}{ S_\mathrm{conf} }

\begin{document}

\title{Liquid Anomalies and Fragility of Supercooled Antimony}

\author{Flavio Giuliani}
\affiliation{Department of Physics, Sapienza University of Rome, 00185 Rome, Italy}
\author{Francesco Guidarelli Mattioli}
\affiliation{Department of Physics, Sapienza University of Rome, 00185 Rome, Italy}
\author{Yuhan Chen}
\affiliation{Department of Physics, Sapienza University of Rome, 00185 Rome, Italy}
\author{Dario Baratella}
\affiliation{Department of Materials Science, University  of Milano-Bicocca, 20125 Milan, Italy}
\author{Daniele Dragoni}
\altaffiliation[Present address:]{Leonardo S. p. A., 00195 Rome, Italy}
\affiliation{Department of Materials Science, University  of Milano-Bicocca, 20125 Milan, Italy}
\author{Marco Bernasconi}
\affiliation{Department of Materials Science, University  of Milano-Bicocca, 20125 Milan, Italy}
\author{John Russo}
\affiliation{Department of Physics, Sapienza University of Rome, 00185 Rome, Italy}
\author{Lilia Boeri}
\affiliation{Department of Physics, Sapienza University of Rome, 00185 Rome, Italy}
\author{Riccardo Mazzarello}
\email[Corresponding author: ]{riccardo.mazzarello@uniroma1.it}
\affiliation{Department of Physics, Sapienza University of Rome, 00185 Rome, Italy}

\date{This manuscript was compiled on \today}
\doi{10.48550/arXiv.2510.25920}


\begin{abstract}
Phase-change materials (PCMs) based on group IV, V, and VI elements, such as Ge, Sb, and Te, exhibit distinctive liquid-state features, including thermodynamic anomalies and unusual dynamical properties, which are believed to play a key role in their fast and reversible crystallization behavior. Antimony (Sb), a monoatomic PCM with ultrafast switching capabilities, stands out as the only elemental member of this group for which the properties of the liquid and supercooled states have so far remained unknown. In this work, we use large-scale molecular dynamics simulations with a neural network potential trained on first-principles data to investigate the liquid, supercooled, and amorphous phases of Sb across a broad pressure–temperature range.
We uncover clear signatures of anomalous behavior, including a density maximum and non-monotonic thermodynamic response functions, which are described by a two-state model based on the structural evolution of the liquid.
Moreover, extrapolation of the viscosity to the glass transition, based on configurational and excess entropies, indicates that Sb is a highly fragile material.
Our results present a compelling new case for the connection between the liquid-state properties of phase-change materials and their unique ability to combine high amorphous-phase stability with ultrafast crystallization.
\end{abstract}


\maketitle

Phase-change materials (PCMs) are employed in optical and electronic storage devices and are interesting candidates for neuromorphic computing \cite{wuttig_phase-change_2007, zhang_phase-change_2019, zhang_designing_2019}. PCMs show a high-conductivity crystalline state (bit ``1") and a metastable low-conductivity amorphous state (bit ``0"). Switching between the two states is reversible, heat-mediated and takes place on nanosecond timescale. These features make it possible to write or process information at the nanoscale through a series of current (or light) pulses that locally amorphize or crystallize the PCM sample.

Common PCMs are alloys of germanium (Ge), antimony (Sb) and tellurium (Te). Monoatomic PCMs made of pure Sb are under active investigation \cite{salinga_monatomic_2018, jiao_monatomic_2020}, as they avoid the segregation issues that affect the alloys \cite{kim_direct_2009, zalden_atomic_2010}. However, Sb undergoes fast exotermic crystallization even near room temperature, hindering the formation of an amorphous phase. 
A promising strategy to enhance the amorphous stability window of pure Sb and other PCMs is nanoconfinement in ultrathin films \cite{kaiser_crystallization_1984, raoux_crystallization_2008, raoux_influence_2009, simpson_toward_2010, chen_size-dependent_2016, salinga_monatomic_2018, jiao_monatomic_2020, dragoni_mechanism_2021, shen_surface_2023, chen_rao_2023}.

Two key features of PCMs, namely the amorphous stability at working temperature and the fast crystallization at slightly higher temperatures, are determined by the glass transition temperature $T_g$ and the fragility of the supercooled liquid phase $m$. 

In PCMs $T_g$ should be higher than ambient temperature to ensure the long-term retention of the amorphous state.
The fragility $m$ is relevant for PCM design, in that a high value of $m$ ensures pronounced atomic mobility upon moderate heating above $T_g$ \cite{zhang_how_2014}. Indeed, high values of $m$ have been reported for Ge$_2$Sb$_2$Te$_5$ ($m\sim90$) \cite{orava_fragility_2012} and GeTe ($m\sim100\div130$) \cite{sosso_breakdown_2012, chen_unraveling_2017, chen_atomistic_2024}. 
However, determining the fragility of PCMs experimentally is challenging owing to the fast crystallization from the supercooled phase. To our knowledge, no experimental data on the fragility of Sb near the glass transition are available in the literature.

The computational cost of ab initio molecular dynamics (MD) based on Density Functional Theory (DFT) is very high, thereby imposing strong finite-size and -time effects that hinder access to the metastable liquid regime at low temperature. 

In recent years, these limitations have been overcome through the development of machine-learned (ML) interaction potentials trained on ab-initio datasets (AIMLP), which enable accurate simulations of systems containing thousands of atoms over tens of nanoseconds. AIMLPs have been successfully employed to explore complex phenomena such as the liquid–liquid critical point in water \cite{gartner_signatures_2020, gartner_liquid-liquid_2022, sciortino_constraints_2025}, and to study crystal growth and dynamical properties
of PCMs \cite{sosso_harnessing_2019, zhou_device-scale_2023, baratella_ge-rich_2025, chen_atomistic_2024, abou_el_kheir_million-atom_2025}, including Sb \cite{holle_thesis_2024}.

In this work, we perform molecular dynamics simulations using an AIMLP based on neural networks presented in Ref. \cite{dragoni_mechanism_2021}, which we also extend by including a previously unreported A17 crystal structure in the training dataset. We investigate the liquid, amorphous, and crystalline phases of Sb across a range of temperatures and pressures. This approach enables us to probe its supercooled dynamics, and structural and thermodynamic anomalies with near-quantum-mechanical accuracy.
We are able to artificially stabilize the supercooled liquid phase of Sb by applying negative pressure, thereby accessing a regime that is normally obscured by rapid crystallization. In this regime, we identify the A17 structure.
In the liquid phase, we observe water-like anomalies that hint at a possible liquid–liquid transition (LLT). These anomalies are accompanied by the emergence of transient A17-like local structures, revealed by bond-orientational order parameters. These structural motifs reflect an intrinsic ordering tendency of the liquid rather than the formation of crystalline precursors. 

Liquid–liquid transitions have been associated with a fragile-to-strong transitions (FST) in the supercooled liquid, as reported in water \cite{angell_water_1993, poole_phase_behaviour_1992, debenedetti_second_2020, amann-winkel_liquid-liquid_2023}, and silica \cite{saika-voivod_fragile-strong_2001}.
FST have also been reported in Te-based and Sb-based PCMs with mostly local octahedral order: Ge$_{15}$Te$_{85}$ \cite{wei_phase_2015}, $\mathrm{Ge}_{15}\mathrm{Sb}_{85}$ and $\mathrm{Ag}_{4}\mathrm{In}_{3}\mathrm{Sb}_{67}\mathrm{Te}_{26}$ \cite{zalden_femtosecond_2019}, and they have been associated to Peierls-like distortions \cite{otjacques_dynamics_2009,zalden_femtosecond_2019}. 
To explore this possibility in Sb, we compute the viscosity and primary relaxation time from MD simulations over a wide temperature range.
We find no indication of an FST transition. Instead, Sb remains very fragile across all temperatures investigated.

This paper is organized as follows:
In the first part, we examine the configurational entropy, the primary relaxation time and the viscosity of the supercooled liquid and extrapolate them to the glass transition temperature to estimate the fragility index (Section \ref{sec:fragility}).
In the second part, we study spontaneous crystallization from the supercooled liquid and characterize the properties of the resulting crystalline phases (Section \ref{sec:xtals}).
Finally, we analyze the anomalous behavior of the liquid state and rationalize it via a Two-State model based on the local liquid structure as quantified by bond-orientational order parameters (Section \ref{sec:liquid-anomalies}).
We draw our final considerations in the Conclusions, followed by a description of the Methods. The Appendices provide the essential theoretical background and detailed analyses supporting our main results.

\begin{figure}[htb!]
    \centering
    \includegraphics[width=0.9\linewidth]{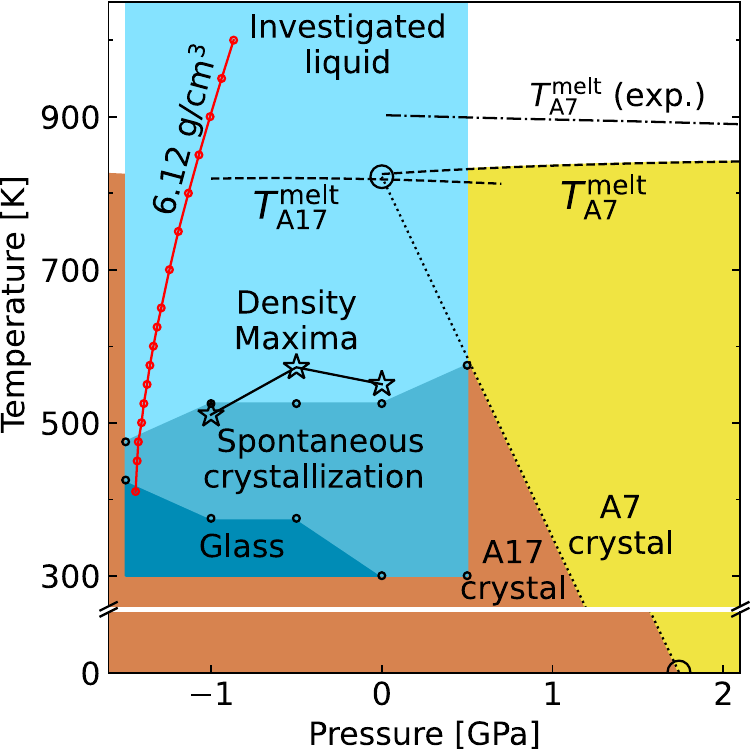}
    \caption{
    $(P,T)$ phase diagram of Sb obtained with the AIMLP of Ref. \cite{dragoni_mechanism_2021}: The blue area indicates the region where we studied the liquid and amorphous phases, with 0.5 GPa resolved isobaric molecular dynamics. Darker blue areas highlight the regions of spontaneous crystallization and glassy dynamics within 4 ns, with calculated boundaries marked by small black circles. Crystals stability regions are colored in brown for the A17 phase and in yellow for A7. The melting curves are determined by direct coexistence simulations (dashed lines; see Methods section), to be compared with the experimental melting line of A7 (dash-dotted). The A17-A7 boundary (dotted line) is sketched between the crossing of melting curves and the enthalpy crossover at $T=0$ (large circles). The existence of a locus of density maxima in the liquid (stars) is a fingerprint of anomalous behavior. For subsequent MD simulations at constant volume, we choose an isochore belonging to the lowest studied pressures, where crystallization is mostly suppressed ($\rho_m=6.12$ g/cm$^3$, red line with small circles).
    }
    \label{fig:PT diagram}
\end{figure}

\section{Fragility of the supercooled liquid}\label{sec:fragility}

\begin{figure}[!t]
	\centering
	\includegraphics[width=.95\columnwidth]{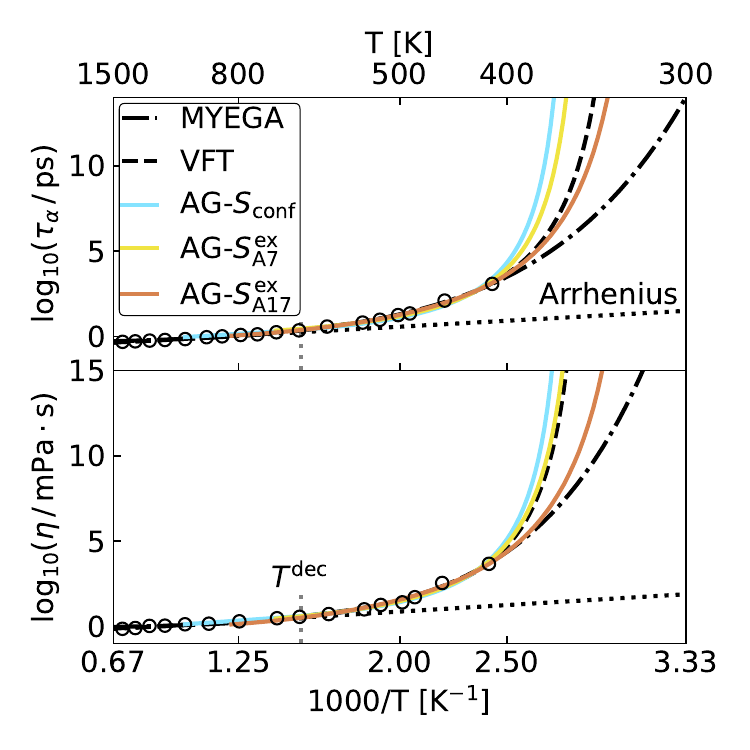}
	\caption{Angell plots showing the outcome of all the fitting methods for the temperature dependence of primary relaxation time $\tau_\alpha$ (top) and viscosity $\eta$ (bottom). The decoupling temperature $T^\mathrm{dec}$ described in the text is marked with vertical lines. Arrhenius fit to the data above 850 K gives an activation energy of 698.4 K for $\tau_\alpha$ and 763.5 K for $\eta$.}
	\label{fig:fragility}
\end{figure}

For PCM applications, it is fundamental to determine the temperature dependence of the viscosity and the glass transition temperature of Sb, as well as to assess the possible presence of a fragile-to-strong transition (FST), which has been linked to enhanced glass stability below the transition temperature and to high crystallization rates above it.
Supercooled liquids are classified as \emph{strong} or \emph{fragile} according to Angell's plot, which depicts the logarithm of the shear viscosity $\eta$, or, alternatively, of the primary relaxation time $\tau_\alpha$, as a function of the inverse temperature normalized by the glass transition temperature $T_g$. The latter is conventionally defined as the temperature at which $\eta(T_g)=\eta^g=10^{12}~$Pa$\cdot$s, or $\tau_\alpha(T_g)=\tau_\alpha^g=100~$s.
Whether a given material is strong or fragile, is determined by the fragility index $m$, which is defined as the slope of the Angell plot at the glass transition: 
\begin{equation}
	m = [\partial\log_{10}\eta / \partial (T_g/T)]_{T=T_g}.
\end{equation}
Strong materials exhibit near-Arrhenius behavior: $\eta \propto  e^{E_a/k_B T}$ with a constant activation energy $E_a$, i.e. a constant slope in the Angell plot. They are typically tetrahedral liquids with strong covalent bonds, such as BeF$_2$, GeO$_2$, SiO$_2$. Since the infinite-temperature limit of viscosity is $\eta_\infty\sim10^{-5}\div10^{-3}~$Pa$\cdot$s for most materials \cite{scopigno_is_2003, mauro_viscosity_2009}, the theoretical lower bound for $m$ is $15\div17$ for strong liquids. 
Fragile materials, on the other hand, have a temperature-dependent activation energy $E_a(T)$, which increases at low temperatures, i.e. a convex curve in the Angell plot. Their fragility index is higher, the steeper is the viscosity increase close to $T_g$. 
In fragile molecular liquids $m$ ranges from $70\div80$ for meta-toluidine, glycerol and ortho-terphenyl to 160 for triphenyl-phosphate.

We perform a preliminary scan of the ($P,T$) phase diagram of pure Sb to determine the optimal range of values for the investigation of the supercooled liquid. We employ the neural network potential presented in Ref. \cite{dragoni_mechanism_2021}. We carry out 4-ns-long simulations of crystallization from the supercooled liquid across a pressure range from –1.5 to 0.5 GPa and temperatures between 300 and 1500 K (blue region in Fig.~\ref{fig:PT diagram}). Our goal is to identify a region where the relaxation time of the supercooled liquid is shorter than the simulation time, and crystallization is sufficiently delayed to allow equilibration of the liquid.
From our preliminary scan, we identify a particularly favorable isochore at a mass density of $\rho_m = 6.12~$g/cm³ (red line in Fig.~\ref{fig:PT diagram}), which is situated in the 
region at negative pressures where the supercooled liquid exhibits its widest stability range in temperature. Along this isochore, crystallization is almost entirely suppressed above $\sim 500$ K, while glass formation only occurs below $\sim 400$ K.
This makes the isochore ideal to probe the deeply supercooled regime. For comparison, the liquid density at the theoretical melting point of 845 K is 6.45 g/cm³ \cite{dragoni_mechanism_2021}, in fair agreement with the value of 6.49 g/cm³ measured at the experimental melting point of 903 K \cite{crawley_density_1972}; the density of
the crystal at 300 K and 1 bar is 6.69 g/cm$^3$ \cite{barrett_crystal_1963}.

Along the $\rho_m = 6.12\,\mathrm{g/cm^3}$ isocore, we directly calculate the viscosity $\eta$ and the primary relaxation time $\tau_\alpha$ from the MD trajectories~\cite{gallo2024supercooled}. 
The viscosity is derived from the off-diagonal components of the pressure tensor using the Helfand-Einstein relation, while $\tau_\alpha$ is extracted from the intermediate scattering function, computed at the $q^*$ value corresponding to the main peak of the static structure factor, that is $2\pi/q^*\approx 3~\Angstrom$. Details of the calculations can be found in Appendix~C of the SI. We find that a linear relation $\eta = G \tau_\alpha$ with $G=1.7\text{ GPa}$ holds above 500 K (see Fig.~S10); small deviations occur at lower temperatures.  
This is not unexpected, since an exact proportionality holds only for $\tau_\alpha(q\to0)$, not for $\tau(q^*)$; the deviations from linearity are a sign that at low temperature collective relaxations over length scales larger than the interatomic distance dominate the dynamics.

We fit the temperature dependence of the viscosity using five functional forms with different levels of approximations, all based on the Adam-Gibbs (AG) theory \cite{adam_temperature_1965}
: i) pure AG, using the liquid's configurational entropy $\Sconf$ computed from the MD simulations; ii) AG using the liquid's excess entropy relative to the A17 crystal ($S^\mathrm{ex}_\mathrm{A17}$) as an approximation of $\Sconf$; iii) AG using the liquid's excess entropy relative to the A7 phase ($S^\mathrm{ex}_\mathrm{A7}$); iv) the Mauro-Yue-Ellison-Gupta-Allan (MYEGA) model \cite{mauro_viscosity_2009}; v) the Vogel-Fulcher-Tamman (VFT) empirical form \cite{scherer_editorial_1992}.
The results of the five fitting methods, together with the extrapolation of the viscosity to low temperatures, are shown in Fig.\ref{fig:fragility} and summarized in Tab.~S4. 
All methods indicate fragile behavior, with no evidence of a FST. 

In the following we discuss the five models in more detail. The AG theory relates the relaxation time of supercooled liquids to the size of cooperatively rearranging regions, which in turn depends on the configurational entropy $\Sconf$:
\begin{equation}\label{eq:AG-Sconf}
	\eta^\mathrm{AG}(T) = \eta_\infty \, e^{\frac{C}{T \Sconf(T)}  }.
\end{equation}

\begin{figure}[hbt!]
	\centering
	\includegraphics[width=\columnwidth]{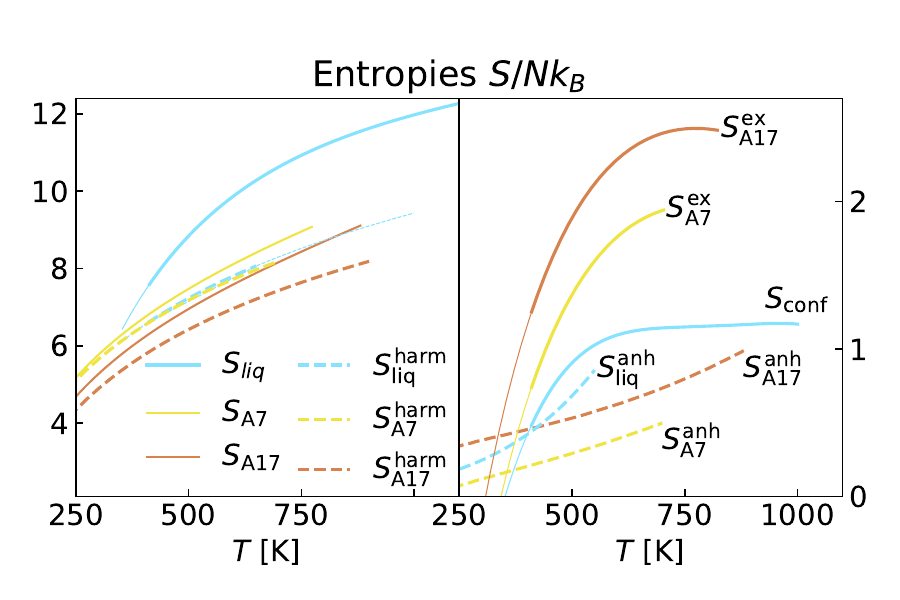}
	\caption{Entropies of the supercooled liquid (blue), of the A7 crystal (yellow) and of the A17 crystal (brown). Left panel: total (solid lines) and harmonic (dashed lines) contributions; the harmonic entropy is evaluated from the phonon spectrum and fitted with a logarithmic temperature dependence. Right panel: configurational and excess (solid) and anharmonic (dashed) contributions. Thinner lines denote the extrapolation outside the data range. See Methods and Appendix~E of the SI for details.}
	\label{fig:entropy_liq_xtals}
\end{figure}
We compute the configurational entropy following the Potential Energy Landscape (PEL) formalism described in Appendix~E of the SI.
In short, we: i) calculate the total entropy via Thermodynamic Integration from a reference liquid, followed by integration of the specific heat; ii) determine the decoupling temperature $T^\mathrm{dec}$, below which the configurational and vibrational degrees of freedom are independent; iii) compute the harmonic entropy for $T<T^\mathrm{dec}$ from the eigenfrequencies of the local minima of the PEL (the so called Inherent Structures, ISs); iv) extrapolate the anharmonic vibrational energy down to $T=0$ and integrate its specific heat to obtain the anharmonic entropy; v) subtract the harmonic and anharmonic vibrational contributions from the total entropy, thus obtaining $\Sconf$ at a single temperature below $T^\mathrm{dec}$; vi) fit the IS energies and integrate their specific heat to determine the configurational entropy as a function of temperature.
As shown in Fig.\ref{fig:entropy_liq_xtals}, we do not observe signs of a FTS in $\Sconf$. The resulting viscosity curve in Fig.\ref{fig:fragility} gives a high $T_g=370$ K and an extremely high value of $m=330$.

In experiments, the {\em excess} entropy $S^\mathrm{ex}$, defined as the difference between the liquid and crystal total entropies, is often used as an approximation of $\Sconf$, by assuming that the vibrational entropies of the liquid ISs and of the crystal are similar. We directly test this assumption by computing the excess entropies $S^\mathrm{ex}_\mathrm{A17}$ and $S^\mathrm{ex}_\mathrm{A7}$ with respect to the A17 and A7 crystals (the properties of these two phases will be discussed in the next section). Direct comparison of these entropies and the configurational entropy in Fig.\ref{fig:entropy_liq_xtals} shows that this assumption is not valid for antimony: the temperature dependence is different, and the absolute values differ by up to a factor of two. Regarding the viscosity fits using the excess entropy, we find that the extrapolated glass transition is quite close to the one of $\Sconf$ if $S^\mathrm{ex}_\mathrm{A7}$ is employed. Instead, using $S^\mathrm{ex}_\mathrm{A17}$ gives different values of $T_g\approx340$ K and $m\approx175$.

Starting from Eq.\ref{eq:AG-Sconf}, the MYEGA model uses constraint theory to relate the configurational entropy to the topological degrees of freedom of the atoms, and then assumes a simple two-state model for the network constraints, which can be either intact or broken. The resulting entropy $\Sconf\propto e^{-1/T}$ goes to zero only at zero temperature, thus ruling out a finite Kauzmann temperature $T_K$:
\begin{equation}\label{eq:MYEGA}
	\eta^\mathrm{MYEGA}(T) = \eta_\infty \, e^{\ln(10)~ a \frac{T_g}{T} \exp\left[ \left(\frac{m}{a}-1\right) \left(\frac{T_g}{T} -1 \right) \right]
	}
\end{equation}
with $a=\log_{10}\left(\eta^g/\eta_\infty\right)$. The MYEGA fit of our viscosity data predicts low values of $T_g=300\div324$ K and $m=74\div95$.

Lastly, the empirical VFT equation \cite{scherer_editorial_1992} is equivalent to assuming $\Sconf\propto 1-T_0/T$ in the AG relation:
\begin{equation}\label{eq:VFT}
	\eta^\mathrm{VFT}(T) = \eta_\infty \, e^{
		\frac{B}{T-T_0}
	}.
\end{equation}
Although this model has severe limitations, it is important because it is the simplest model bearing a positive Kauzmann temperature $T_K\sim T_0$. The VFT fit gives intermediate values of $T_g$ and $m$ between those obtained from the AG and MYEGA models.

In summary, due to the inevitable extrapolation errors below 410 K, our best estimate is $T_g\in[300,370]$ K and $m\in[74,330]$, with the highest and lowest values coming from the AG relation and the MYEGA fit, respectively. Replacing the configurational entropy with the A7 or A17 excess entropies in the AG relation gives results that are closer to the empirical fits.

We conclude that Sb is a highly fragile material. No evidence of a transition to strong behavior is found along the isochore $\rho_m = 6.12\,\mathrm{g/cm^3}$ for $T \geq 410$ K, before the liquids falls out of equilibrium within our simulation range.
In the next section, we focus on the crystallization properties of the system at the same density.


\section{Crystallization and crystal phases}\label{sec:xtals}

\begin{figure}
    \centering
    \includegraphics[width=\linewidth]{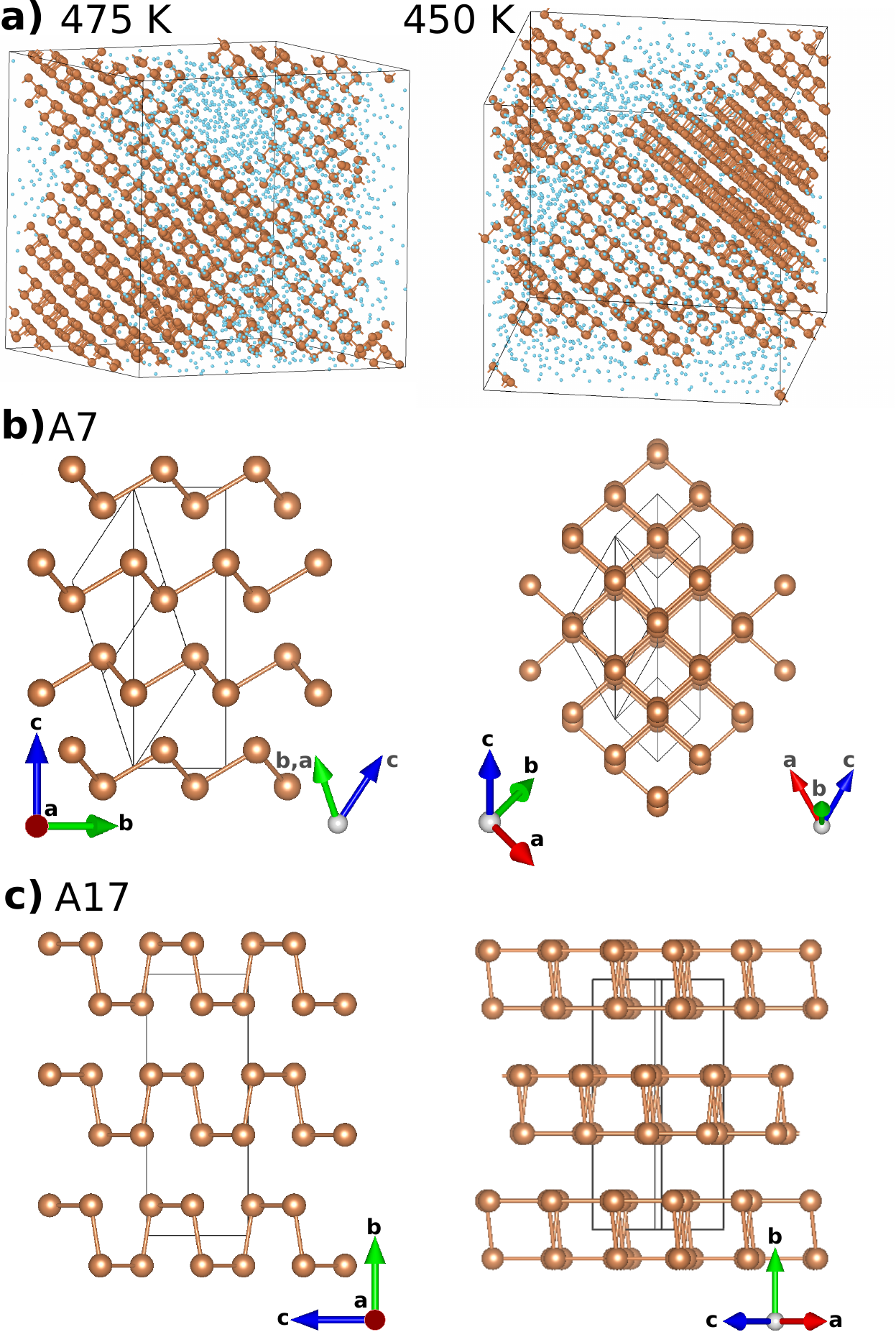} 
    \caption{Crystallization properties and stable crystal phases of antimony.
    a) Along the isochore $\rho_m = 6.12~$g/cm³ spontaneous crystallization from the supercooled liquid occurs at temperatures of 475 K (left) and 450 K (right). The two snapshots of the partially crystallized samples show the formation of one or more nuclei of A17 in the [101] view (see the right panel in c). Liquid particles are visualized in light blue color with a smaller radius; crystal particles belonging to the main cluster(s) are colored in brown with larger radius and are connected by bonds.
    b-c) Atomic structure of A7 and A17 crystals in the conventional cells. We use different colors and decreasing thickness to represent the bonds with the first (brown), second (green) and third (blue, only for A17) groups of nearest neighbors (NNs).
    b) A7: side view slightly tilted from the [100] direction, highlighting the ABC stacking of $\beta$-antimonene pseudo-bilayers; a semitransparent polyhedron represents the distorted octahedral environment with the six NNs.
    c) A17: side views slightly tilted from the [100] direction, showing the "washboard" shape of the bilayers, and from [101], highlighting the AB stacking of symmetric-$\alpha$-antimonene and octahedral-like patterns; a semitransparent polyhedron represents the distorted defective octahedral environment with the five NNs.
    }
    \label{fig:crystallization and crystals}
\end{figure}

Bulk antimony undergoes fast crystallization from the glassy phase at ambient conditions, so fast that, to our knowledge, no experimental data exist on the crystallization time of the bulk material, but only for ultra-thin samples where the amorphous phase can be stabilized. In a 5 nm film \cite{salinga_monatomic_2018}, the crystal incubation time varies from 100 s at 300 K to $\approx 3\cdot 10^{-4}$ s at 400 K following an Arrhenius law; for a 10 nm film, extrapolation gives $\approx 10^{-4}$ s at 300 K and $\approx 10^{-6}$ s at 400 K. Ab initio MD simulations have shown that the crystal incubation time depends on the preparation protocol, on the density of the bulk sample and on finite-size effects \cite{salinga_monatomic_2018}. More specifically, increasing the quenching rate from 9.5 K/ps to 300 K/ps led to an increase of the crystal incubation time from 100 ps to 500 ps, in a model of 360 atoms at 300 K. Reducing the density by 7\% gained a seven-fold increase, while doubling the number of atoms resulted in a six-fold increase of the incubation time. Recent works using AIMLPs scaled up to thousands of atoms and confirmed the importance of density in the incubation time \cite{dragoni_mechanism_2021, holle_importance_2025}. After the incubation time, crystallization from the metastable liquid is dominated by the growth of supercritical nuclei \cite{dragoni_mechanism_2021}. In fact, the velocity of the crystal front in the bulk -measured at the liquid density at the melting point, using the same AIMLP of our paper- varied from 3 m/s at 300 K to a maximum of 35 m/s at 600 K \cite{dragoni_mechanism_2021}.

Despite working at negative pressures, crystallization occurs on the nanosecond timescale (Fig.\ref{fig:crystallization and crystals}a) in the range $400-500$ K.
In the same temperature range, the primary relaxation time $\tau_\alpha$ (defined in Appendix~C of the SI) increases from $\sim 0.01$ ns to $\sim 1$ ns. Equilibrating and sampling the system becomes challenging when $\tau_\alpha$ approaches the crystallization time.
In this regime, often referred to as "no man’s land" in the $(P,T)$ diagram,
it is physically impossible to equilibrate the supercooled liquid.

To better understand the mechanism of crystallization,
we analyze the structural properties of the recrystallized models. Snapshots from two trajectories at 450 K and 475 K along the selected isochore are shown in Fig.\ref{fig:crystallization and crystals}a, where particles are classified as liquid (in blue, with smaller radius) or crystalline (in brown, with larger radius) according to the bond-order parameter $q_4^\mathrm{dot}$ (see Methods).

Surprisingly, we find that, at negative pressure, Sb does not crystallize into the ambient-pressure A7 structure, but rather into an A17 “black phosphorus”–type phase. We stress that the A17 structure was not included in the training set of \cite{dragoni_mechanism_2021}. This unexpected result suggests that, at negative pressures, Sb favors a layered polymorph not previously reported in its bulk phase diagram. Traces of the structural order of the A17 phase appear in the local structural motifs of the supercooled liquid, as will be discussed in Section~\ref{sec:liquid-anomalies}.

The A7 and A17 crystal structures are shown in Fig.~\ref{fig:crystallization and crystals}b-c.
A7 is a rhombohedral phase with space group $R\overline{3}m$, which can be derived from a simple cubic lattice via a trigonal distortion along the [111] direction. Alternatively, it can be viewed as a pseudo-layered ABC stacking of hexagonal $\beta$-antimonene sheets, with significant interlayer chemical bonding \cite{akturk_single-layer_2015, xue_recent_2021}.
A17, on the other hand, has space group $Cmce$ and consists of a pseudo-layered AB stacking of symmetric $\alpha$-antimonene.
Both structures exhibit strong Peierls distortions, characterized by an alternation of short and long bonds along the direction of approximately collinear $p$ orbitals.
In the A7 phase, the distortion takes place between the three shorter intra-bilayer bonds and the three inter-bilayer bonds with the second nearest neighbors. In the A17 phase, the distortion takes place between two shorter and two longer bonds parallel to the layer plane; the intra-bilayer short bond approximately perpendicular to the layer plane is not associated with a distortion, because the angle with the sixth nearest neighbor on the next bilayer is far from flat, being $120\degree$-$145\degree$.
We present a detailed quantitative analysis of the distortions through the Angular-Limited Three-Body Correlation (defined in Appendix~A of the SI) in Section \ref{sec:liquid-anomalies}.

We also note that these results shed light on the simulations based on the same AIMLP of Ref.~\cite{Ritarossi2025PSS-RRL}, where it was reported that thin layers of Sb confined in an artificial superlattice crystallize into a layered phase that is different from the A7 crystal.

We compare the enthalpy differences between the two crystal phases computed with the NN potential and with DFT (Fig.~S4). DFT simulations confirm that the A17 phase becomes stable at sufficiently negative pressures, with a transition pressure of $P^\mathrm{A17-A7}_{\mathrm{DFT}}=-1.16~$GPa. Hence, DFT predicts the transition to occur at lower pressures than the AIMLP: $P^\mathrm{A17-A7}_\mathrm{AIMLP} - P^\mathrm{A17-A7}_{\mathrm{DFT}} = 2.9$~GPa. 

These observations motivated us to extend the AIMLP of Ref.~\cite{dragoni_mechanism_2021} by incorporating A17 crystal configurations into its training dataset. The new AIMLP was developed within the MACE framework \cite{MACE}.  
It provides a more accurate description of the relative enthalpies and densities of the two phases across a wide pressure range compared to the original potential (Fig.~S5). 

In the next section, we employ the new AIMLP and consider state points in an extended region of the phase diagram, where the liquid shows anomalous thermodynamic behavior. We demonstrate that traces of the A17 structure can be found in the supercooled liquid state as well.

\section{Liquid anomalies}\label{sec:liquid-anomalies}

\begin{figure}
	\centering
	\includegraphics[width=\columnwidth]{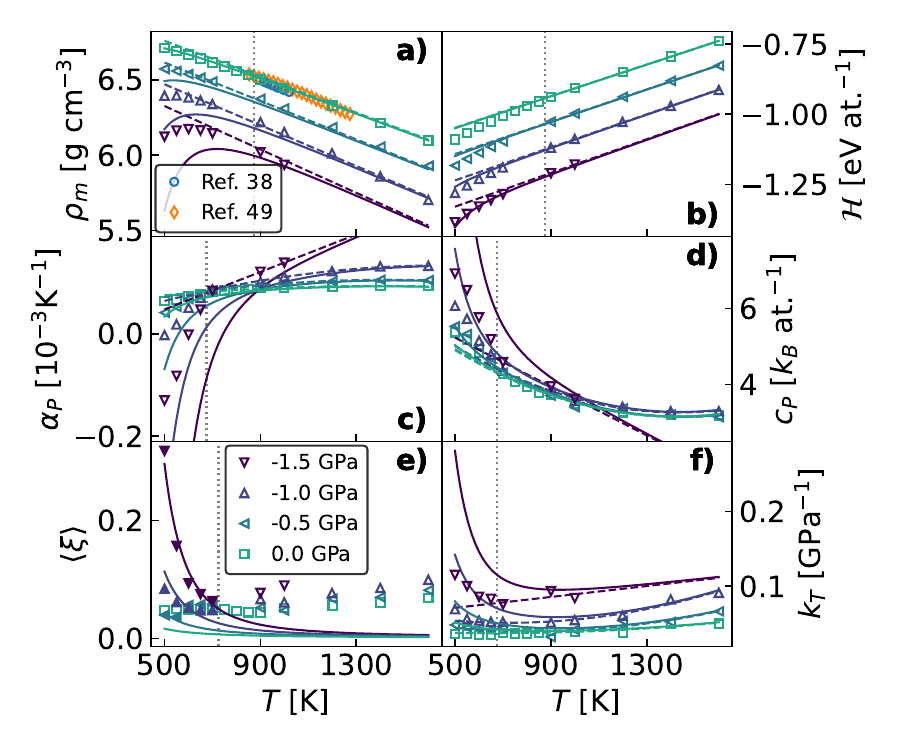} 
	\caption{
		Anomalies in thermodynamic response functions and in structural data (markers) derived from the MD simulations using the MACE potential, and two-states (TS) fit (solid lines) using Eq. \ref{eq:two-states_S_highT} and Eqs. \ref{eq:two-states_volume}-\ref{eq:two-states_kT}.
		a) Mass density $\rho_m$ and
		b) enthalpy $\mathcal{H}$.
		c) Thermal expansion coefficient $\alpha_P=V^{-1} (\partial V / \partial T)_P$ and
		d) isobaric heat capacity $c_P=N^{-1}(\partial \mathcal{H}/\partial T)_P$, both directly computed from the data in the top panels.
		e) Equilibrium fraction $\langle \lowTpopulation \rangle$ of the low-temperature "anomalous" liquid and
		f) Isothermal compressibility $k_T=-V^{-1} (\partial V / \partial P)_T$.
		The dashed lines in panels a-d, f are the "background" high-temperature regime defined in Eq. \ref{eq:two-states_anomalyX}; The backgrounds of panels a and b are fitted over temperatures $T \geq T_1^B = 900$ K; the backgrounds of response functions in panels c, d and f are fitted over temperatures $T \geq T_2^B = 700$ K; $ T_1^B$ and $ T_2^B$ are shown as vertical dotted lines.
		The equilibrium fraction $\langle \lowTpopulation \rangle$ is a structural measure of the anomaly, equal to the relative area of the secondary population in the distribution of the order parameter $q_3$; the fit with Eq. \ref{eq:two-states_S_highT} is carried on temperatures $T \leq T_\xi = 700$ K, which is shown as a vertical dotted line, and on pressures $P \leq P_\lowTpopulation=-0.5$ GPa, because the extraction of $\langle \lowTpopulation \rangle$ from $q_3$ is less noisy in this region (filled markers).
        Panel a also includes experimental data about the density at ambient pressure \cite{crawley_density_1972, tsuchiya_compressibility_1997},  which are in excellent agreement with the MACE values.
	}
	\label{fig:anomaly isobars}
\end{figure}

\begin{figure*}[hbt!]
	\centering
	\includegraphics[width=\linewidth]{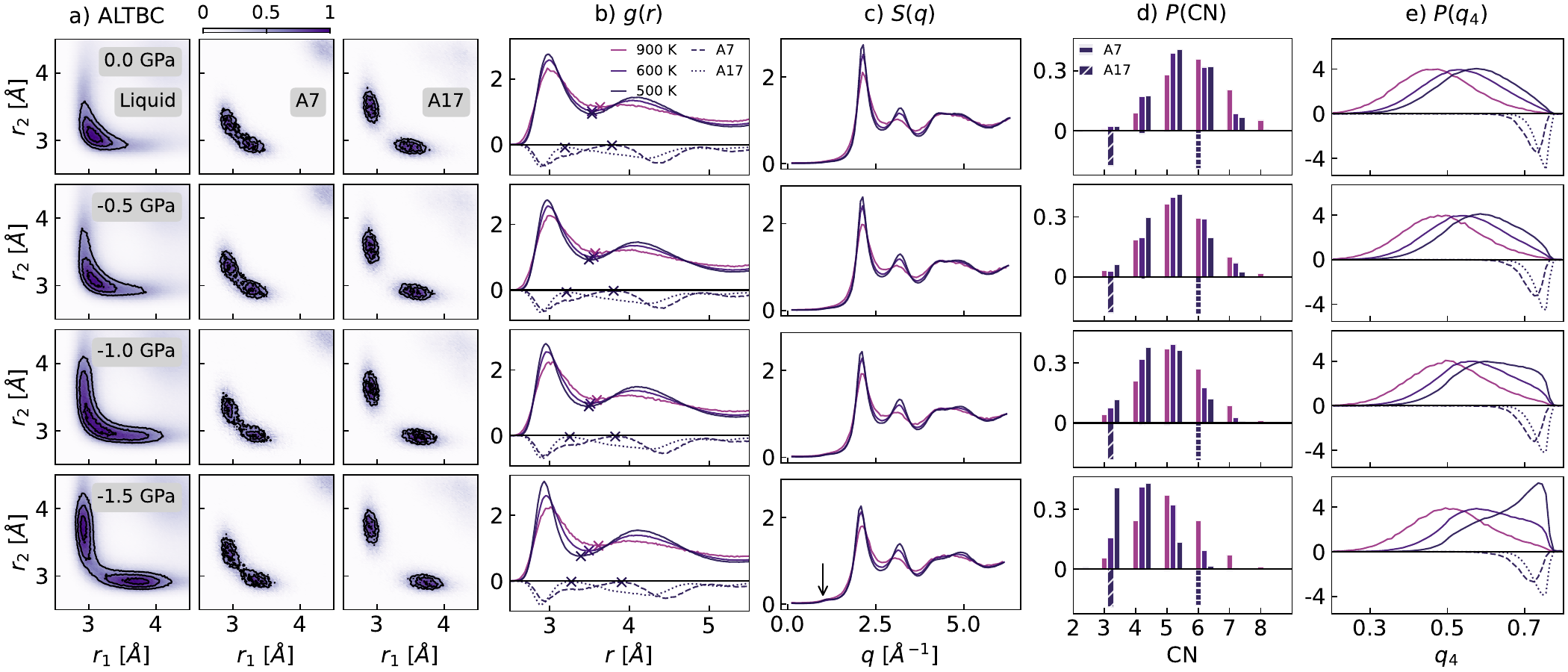} 
	\caption{
		Structural analysis of liquid and crystal phases. Angular-limited three-body correlation function (ALTBC), radial pair distribution $g(r)$, static structure factor $S(q)$, coordination number (CN), and Steinhardt order parameter $q_4$, of the liquid phase (at 900 K, 600 K and 500 K) and of the crystal phases ("A7" and "A17" at 500 K) at four different pressures. Contour lines at 0.5, 0.7 and 0.9 are drawn in the ALTBC plots. For each state point the radial cutoffs for CN and $q_4$ is set to the first local minimum of the corresponding $g(r)$ and is shown as crosses in the $g(r)$ panel. Crystal data in the $g(r)$, CN and $q_4$ panels are shown as negative-valued curves and scaled by 1/5. The pre-peak in $S(q)$ is marked with an arrow at the lowest pressure.
    }
	\label{fig:isotherms and structure}
\end{figure*}

\begin{figure}[htb!]
	\centering
	\includegraphics[width=.9\columnwidth]{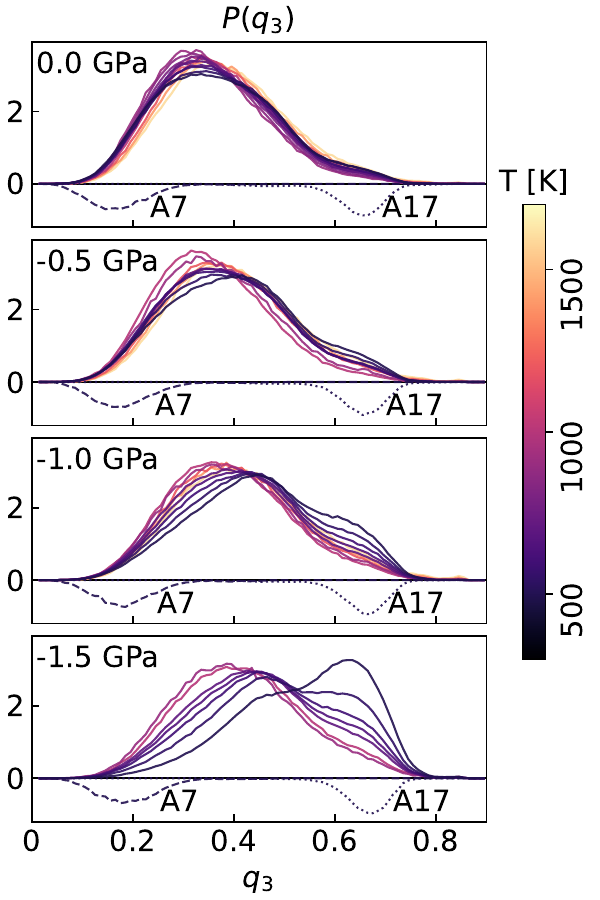}
	\caption{   
		Low-temperature anomalies in the probability distribution of the bond orientational order parameter $q_3$ in the liquid and supercooled phases, from 1600 K to 500 K (solid lines).  The reference distributions in the two crystal phases at 500~K are shown as negative-valued curves (A7: dashed, A17: dotted) and scaled by 1/10. The
        results indicate that i) at low temperature the liquid becomes more ordered, and ii) a secondary type of liquid with high $q_3$ emerges. The local angular structure of this secondary liquid is similar to that of A17 rather than A7.}
	\label{fig:q3 liquid} 
\end{figure}

Having established the crystallization behavior and the emergence of the A17 structure at negative pressures, we perform 
isothermal-isobaric MD simulations of liquid Sb in the region between $-1.5$ GPa and 0.5 GPa and between 1500 K and 300 K.
We observe unexpected structural and thermodynamic anomalies, reminiscent of those found in water and other tetrahedral systems. In the following we present the results obtained with the extended AIMLP described in the previous section; the old potential yields similar results, as shown in the supplement. 

Fig.~\ref{fig:anomaly isobars} shows that, at negative pressures below -1 GPa, the mass density $\rho_m$ exhibits a maximum at temperatures between 500 and 600 K, while at higher pressures, the maximum shifts to lower temperatures below 500 K, at which, however, the supercooled liquid is not accessible since the relaxation times exceed the simulation times (panel a);
this implies that the thermal expansion coefficient $\alpha_T$ goes to zero and even becomes negative at -1 GPa or below, as shown in panel c.

The isobaric heat capacity $c_P$, derived from the enthalpy shown in panel b, shows an anomalous increase (panel d), indicating an enhancement in fluctuations.
However, the lines of heat capacity maxima always lie within the region of slow relaxation ($T \lesssim 500 K$).
The resulting trends provide compelling evidence for the presence of liquid anomalies, and suggest the existence of
a liquid-liquid transition~\cite{russo_physics_2022}.

In tetrahedral systems the thermodynamic anomalies are linked to underlying changes in local structure. In the following we thus perform a detailed analysis of short-range order at different pressures and temperatures. The results, reported in Fig.~\ref{fig:isotherms and structure}, indicate that at low temperature and low density the liquid is more ordered, locally less dense and has more pronounced Peierls-like distortion.
In fact, we observe an enhancement of the first and second peaks of the radial pair distribution, located at $3.0 ~\Angstrom$ and $4.1~\Angstrom$, respectively, accompanied by a reduction of the minimum in-between. 
The average number of neighbors decreases from 5 at 900 K to 4 at 300 K, indicating lower local density.
The angular-limited three-body correlation function (ALTBC) shows that symmetric collinear bonds with $r_1\sim r_2\sim 3.2~\Angstrom$ are depleted, while Peierls-distorted collinear bonds with $r_1\sim2.9~\Angstrom ,~ r_2\sim 3.7~\Angstrom$ become more frequent; this is confirmed by the formation of a pre-peak in the static structure factor at $\sim 1.16~\Angstrom^{-1}$ ($5.42~\Angstrom$ period in real space), approximately at half wavevector of the main peak.
Furthermore, the angles between nearest neighbors approach the values found in the two crystal phases (either A17 or A7), i.e. $\sim 90\degree$ and $\sim 165\degree$ (see Fig.~S3).

These structural features combined with the thermodynamic anomalies suggest that monoatomic Sb behaves similarly to H$_2$O and Si \cite{gallo_water_2016, russo_physics_2022} and can potentially exhibit a liquid–liquid transition between a high-density disordered liquid and a low-density, locally ordered one.
Analogously to water, the anomalous behavior could be explained by the formation of locally favored structures that emerge from the disordered liquid background~\cite{russo2014understanding}.

In order to shed light on the liquid–liquid transition, we characterize the local structure using the Steinhardt bond-orientational order parameter $q_3$ \cite{steinhardt_bond-orientational_1983}, which has previously been shown to be sensitive to changes in local coordination in tetrahedrally bonded liquids such as water \cite{russo2014understanding}. The $q_3$ parameter probes the angular arrangement of nearest neighbors through spherical harmonics of degree $l=3$, and is particularly sensitive to deviations from centrosymmetric local environments, providing a measure of distortions of the octahedral geometry and of local inversion-symmetry breaking. For each atom i, $q_3$ is computed from the orientations of the bonds connecting i to its nearest neighbors, yielding a scalar measure of local bond-orientational order that discriminates between the two liquid states.

Fig.~\ref{fig:q3 liquid} shows the distribution of $q_3$ in the liquid phase as a function of temperature, for decreasing pressure (top to bottom).
The plots reveal the emergence of a two-population distribution at low temperature. During cooling, at all pressures a shoulder appears at $q_3 \sim 0.65$. 
This implies that the population of a liquid with more distorted octahedral-like local angular geometry increases at low temperature. 
Notably, the shoulder position coincides with the main peak of $q_3$ in the A17 crystal rather than with that of A7.
Therefore, the local structure of the low-temperature population is similar to that of  A17.
We find similar indications in the analysis of the bond orientational order parameter $q_4$ \cite{steinhardt_bond-orientational_1983}, shown in Fig.~\ref{fig:isotherms and structure} and further discussed in Appendix~A of the SI. We note that the shoulder cannot be ascribed to the presence of A17 subcritical nuclei, since the fraction of crystal-like particles in our models (computed using the $q_4^\mathrm{dot}$ order parameter, which also includes medium-range order: see Methods section) is always below 1 \%. In the supplement, we further analyze the models using other structural order parameters, including an octahedral order parameter $q_\mathrm{oct}$, which measures the deviation from a perfect octahedral angular geometry.

We denote with $\lowTpopulation$ the fraction of the low-temperature liquid population in a given configuration. We measure its average value at equilibrium as a function of temperature and pressure, $\langle \lowTpopulation \rangle(T,P)$, by decomposing the probability distribution of $q_3$ into two Gaussians and computing the relative area of the secondary Gaussian.
The data for $\langle \lowTpopulation \rangle$ is shown in Fig.~\ref{fig:anomaly isobars}e. Within the sampled metastable liquid region, we always observe values of $\langle \lowTpopulation \rangle$ below 30\%. This is in agreement with the fact that the Widom line, marking the crossover from the high-temperature liquid to the low-temperature one, i.e. $\langle \lowTpopulation \rangle \approx 50\%$, lie inside the inaccessible region of fast crystallization.

We fit these thermodynamic and structural anomalies using the simple Two-State (TS) model introduced by H. Tanaka in 2000 \cite{tanaka_simple_2000}. The two relevant states are i) the high-temperature disordered liquid and ii) the low-temperature liquid with short-range order similar to A17.
At high temperatures, which correspond to low $\lowTpopulation$, the TS model yields:
\begin{equation}\label{eq:two-states_S_highT}
	\langle \lowTpopulation \rangle \sim \exp\left[\beta(\Delta E - T \Delta S + P\Delta V)\right],
\end{equation}
where $\Delta E$, $\Delta S$ and $\Delta V$ are, respectively, the differences of energy, entropy and volume per atom between the high-temperature liquid and the low-temperature liquid (e.g., $\Delta E=E_\mathrm{high-T}-E_\mathrm{low-T}$), $T$ is the temperature, $P$ is the pressure and $\beta=1/k_B T$.
In the simplest approximation, $\Delta E$, $\Delta S$, $\Delta V$ do not depend on temperature nor on pressure, leading to a three-parameter fitting form for all $\langle \lowTpopulation \rangle (T,P)$ data.
This enables a simple description of the anomalies of any thermodynamic observable $X$, by assuming that the "background" behavior due to the high-temperature liquid, $X_B$, is complemented by an anomalous contribution, $\Delta X$, from the low-temperature liquid population. The latter contribution is weighted by $\langle \lowTpopulation \rangle$ and becomes significant at low temperature:
\begin{equation}\label{eq:two-states_anomalyX}
	X(T,P) = X_B(T,P) - \Delta X \cdot \langle \lowTpopulation \rangle .
\end{equation}
Details of the fitting procedure are described in the Methods.

The best fit parameters are $\Delta E / k_B = 1416$ K, $\Delta S / k_B = 6.93$, $\Delta V=-13.2~\Angstrom^3/\text{at.}=-7.98$ cm$^3$/mol. In Table S5 we report the full list of parameters.
The volume difference $\Delta V$ is about 40\% of the typical volume per atom (which spans between 30 and 35 $\Angstrom^3/\text{at.}$ in the range studied).
The entropy difference implies that the ratio between the degeneracy of the high-temperature and the low-temperature structures is $e^{\Delta S/k_B}=1.02 \cdot 10^{3}$. Compared to liquid water \cite{tanaka_simple_2000}, $\Delta V$ is about $1.3$ times smaller and the degeneracy ratio is about $12$ times lower.

In summary, we provide evidence for the existence of two liquid states that differ in the local ordering, similar to what occurs in water~\cite{russo2014understanding} and other tetrahedral materials~\cite{shi2019distinct}. Our analysis shows that the low-temperature liquid population exhibits a local octahedral geometry resembling the A17 crystal rather than the A7 one. We interpret these results within a TS model, which explains the density maxima and the thermodynamic anomalies.

Our analysis also reveals that a FST is likely to occur in deeply supercooled Sb. However, since the population of the low-temperature liquid remains below 30\%, we do not observe a complete transition from the high-temperature to the low-temperature phase within the temperature range accessible to our simulations. 

\section{Conclusions}
Our study addressed the long-standing gap in understanding the liquid and supercooled phases of elemental antimony (Sb), revealing that it shares key features with other group IV–VI phase-change materials.

Using an ab initio machine-learned potential trained on DFT data~\cite{dragoni_mechanism_2021}
we investigated the viscosity of the liquid down to $410$ K along the 6.12 g/cm$^3$ isochore. We directly extracted both the viscosity $\eta$ and the primary relaxation time $\tau_\alpha$ from the atomistic trajectories, and in parallel computed the configurational entropy $\Sconf$ within the potential-energy-landscape formalism. All three quantities exhibit fragile behavior; in particular, the viscosity changes by 2.5 orders of magnitude over a temperature range of 300 K above 410 K. No clear signs of a fragile-to-strong transition are observed within the accessible temperature range. Extrapolation with the Adam-Gibbs relation and other fitting forms yields a glass transition temperature in the range $T_g\in[300,370]$ K and a fragility index $m\in[74,330]$. We note that, in Ref.~\cite{holle_thesis_2024}, a moment-tensor AIMLP was employed to estimate the glass transition temperature of Sb: $T_g=333$ K when extrapolated from moderately supercooled data above 800 K (Fig.~4.15 therein), or $T_g\in[180,320]$ K from the maximum slope of the isochoric specific heat (Fig.~C6 therein).
Based on our results, antimony can be classified as an ultra-fragile material, unless a LLT-driven fragile-to-strong transition around $\sim 400$ K slows down the viscosity divergence. The latter possibility is supported by the estimated small fraction of the low-temperature liquid, and by the similarities with water, which becomes a strong liquid at low temperature~\cite{de2018fragile,shi2018origin}.

While studying crystallization, we found that Sb spontaneously nucleates into the A17 phase from the bulk supercooled liquid at negative pressure. Previous reports on the A17 phase in Sb were limited to ultrathin samples; the only bulk phase considered stable in the literature at low pressure was A7. We verified the existence of a {\em bulk} A17-A7 transition by direct coexistence simulations. Including the A17 structure in the AIMLP training set allowed us to confirm that the A17 phase is stable at negative pressures below –1 GPa.

Finally, we identified water-like anomalies in liquid Sb, including density maxima, anomalous isobaric heat capacity and the emergence of a population of locally more ordered structures at low temperature, identified via the $q_3$ order parameter. Noticeably, the local structure of the anomalous low-temperature liquid resembles that of the A17 crystal.
Since the anomalous population remained below $<30\%$ across the accessible range, we fitted the data with the high temperature expansion of Tanaka's two-state model. Our model accounts for all anomalies satisfactorily and suggests that a liquid–liquid transition is hidden in the "no man's land" of fast crystallization below $\sim 500$ K and negative pressures. Nevertheless, the transition has observable effects at ambient pressure, relevant to phase-change applications. In particular, thermodynamic anomalies persist even at zero pressure, as evidenced by the behavior of the heat capacity and the $q_3$ distribution. Furthermore, a flattening of the density curve is observed at zero pressure, although the maximum lies within the no-man’s land. 

Overall, these findings establish elemental antimony as a valuable model system for investigating the interplay between liquid-state anomalies, structural and dynamical properties, and phase-change functionality.

\section*{Methods}

\subsection*{First principles calculations}
We computed the enthalpy-pressure phase diagram of Sb crystal phases within DFT, employing the Quantum Espresso 7.0 software \cite{QE-2009,QE-2017}. We used the PBE functional, a scalar-relativistic optimized norm-conserving Vanderbilt pseudopotential for Sb with 15 valence electrons and a plane-wave expansion of Kohn–Sham orbitals up to an energy cutoff of 90 Ry. We optimized the Gaussian smearing and the size of the uniform mesh for the Brillouin zone integration at zero pressure by converging the total energy within 1 meV/atom and each diagonal component of the pressure within 0.05 GPa.

\subsection*{Molecular dynamics}
We investigate supercooled liquid Sb by extensive MD simulations with a time step of $2$ fs. We employ two AIMLPs, namely the potential introduced in Ref. \cite{dragoni_mechanism_2021} based on the Behler-Parrinello approach and a newly developed MACE potential \cite{MACE} trained on an extended dataset that incorporates A17 crystal configurations.\\ 
\emph{Behler-Parrinello potential}: The weights of the neural network were optimized on energies and forces from configurations of crystal, liquid and amorphous Sb phases, computed via DFT. A $6.6~\Angstrom$ cutoff was used. For clarity, we report here the DFT details provided in Ref. \cite{dragoni_mechanism_2021}: PBE functional, norm-conserving pseudopotential and plane-wave expansion of Kohn–Sham orbitals up to an energy cutoff of 40 Ry; the Brillouin zone integration was performed over a uniform mesh by keeping approximately the same k-point linear spacing of $0.13~\Angstrom^{-1}$ for all configurations. Following \cite{dragoni_mechanism_2021}, we add a long-range dispersion correction to the AIMLP, using the "D2" semi-empirical scheme by Grimme \cite{grimme_consistent_2010} with a cutoff of $12~\Angstrom$. We apply the same correction in DFT calculations for the crystal phase diagram in Fig.~S4.\\
\emph{MACE potential}: The potential was developed by fine-tuning the MACE mace-omat-0 foundation model \cite{MACE}, pretrained on the omat2024 DFT database \cite{OMAT24}, using the recent multi-head fine-tuning approach \cite{HYDRA} implemented in the MACE code \cite{MACE}. The DFT fine-tuning dataset coincides with that employed for the development of the Behler–Parrinello potential, supplemented with additional configurations of the A17 crystalline phase. The same D2 corrections discussed above are added to the potential. The potential and the database have been uploaded to Materials Cloud (DOI: ...).

We use the LAMMPS software \cite{LAMMPS} (version 11 Aug 2017 when not specified) with the following specific commands for the pair styles: \texttt{runner} for the Behler-Parrinello AIMLP, \texttt{momb} for the vdW correction, and \texttt{hybrid/overlay} to combine them. In the calculation of the dynamical matrices, we use the \texttt{dynamical\_matrix} command introduced in a more recent version of LAMMPS (2 Aug 2023); in this case, we use the equivalent \texttt{hdnnp} pair style command for the Behler-Parrinello AIMLP interface. In isothermal-isochoric (NVT) simulations we use a Bussi-Donadio-Parrinello thermostat (command \texttt{fix temp/csvr}) \cite{bussi_canonical_2007}. In isothermal-isobaric (NPT) simulations we use a Nos\`e-Hoover-chain with chain length of 3 for both temperature and pressure coupling, employing the Martina-Tuckerman-Tobias-Klein decomposition scheme (command \texttt{fix npt}) \cite{tuckerman_liouville-operator_2006}.
We use a cubic supercell with periodic boundary conditions containing 4096 Sb atoms, both in NVT and NPT simulations of the liquid and glassy phases.
The same setting is used for NPT simulations with the MACE potential, with a more recent version of LAMMPS.
The main NVT simulations are carried at $6.12~$g/cm$^3$ mass density (about $33.0~\Angstrom^3$ atomic volume), i.e. inside a box of size $51.3388~\Angstrom$. We quench the system at 9.5 K/ps (when not specified otherwise) and equilibrate at each temperature while monitoring the potential energy and the relaxation dynamics through the Intermediate Scattering Function (ISF) and the Mean Square Displacement (MSD) (defined in Appendix~C of the SI). For temperatures at which the ISF primary relaxation time is $\sim 1~$ns or less, we subsequently run NVE dynamics to measure ISF and MSD at equilibrium.

For the NVT simulation of crystals at $6.12~$g/cm$^3$, we use supercells with a similar number of atoms to the liquid cell: for the A7 phase, 4320 atoms in an hexagonal-prism box with cell vectors $\mathbf{a}=(53.118,0,0)$ $\mathbf{b}=(-26.559,46.002,0)$, $\mathbf{c}=(0,0,58.403)$, in units of $\Angstrom$; for the A17 phase, 4608 atoms in an orthorhombic box with $a_x=50.617~\Angstrom$, $b_y=49.517~\Angstrom$, $c_z=60.727~\Angstrom$. 

\subsection*{Identification of crystal nuclei}
Among all trajectories in the liquid and supercooled phase, we select equilibrium ones by requiring a time window at least twice larger than the ISF primary relaxation time and containing in every frame no more than 40 crystalline atoms (0.98\%) and crystalline nuclei smaller than 30 atoms (0.73\%). We classify particles as crystalline based on the established $q_4^\mathrm{dot}$ bond order parameter \cite{dragoni_mechanism_2021,tanaka_revealing_2019}, which is defined in terms of spherical harmonics with $l=4$. The two radial cutoffs for $q_4^\mathrm{dot}$ are set as equal to the first and second local minima of the $g(r)$ at 900 K at that density (in NVT, or pressure in NPT) and the threshold for crystallinity was set to 0.75. We define a crystal nucleus by grouping crystalline particles that are closer than $r_\mathrm{cl}=3.70~\Angstrom$; the latter value is slightly larger than the first local minimum of the $g(r)$.
We apply the same requirements to the liquid before conjugate-gradient minimization to generate the ISs, and to the ISs themselves.

\subsection*{Two-states model}
We rationalize the thermodynamic and structural anomalies in the supercooled liquid by the Two-State (TS) model. The two states correspond to i) a high-temperature disordered liquid and ii) a low-temperature ordered liquid with local structure similar to A17. We identified the two states by a bimodal distribution of an order parameter representing two different populations of local structures. We chose the Steinhardt bond-orientational order parameter $q_3$.

Let $A_-$ and $A_+$ be the areas under the two Gaussian curves fitted to the distributions of $q_3$ in Fig.\ref{fig:q3 liquid}, where $A_-$ refers to the Gaussian developing at low temperatures and pressures. We estimated the equilibrium fraction of low-temperature structures in the liquid as $\langle \lowTpopulation \rangle = A_-/(A_- + A_+)$. Since nucleation is very fast at low temperature, we were able to study only the TS model in the low-$\langle \lowTpopulation \rangle$ (high-temperature) region: $\langle \lowTpopulation \rangle < 30\%$.

Explicitly, Eq.\ref{eq:two-states_anomalyX} takes the following form for volume, mass density and enthalpy:
\begin{align}
    V(T,P) &= V_B(T,P) - \Delta V \cdot \langle \lowTpopulation \rangle , \label{eq:two-states_volume}\\
    \rho_m(T,P) &\approx \rho_m^B(T,P) \left( 1 + \frac{\Delta V}{V} \langle \lowTpopulation \rangle \right), \label{eq:two-states_density}\\
    \mathcal{H}(T,P) &= \mathcal{H}_B(T,P) - (\Delta E + P \Delta V) \cdot \langle \lowTpopulation \rangle , \label{eq:two-states_enthalpy}
\end{align}
where each of the quantities corresponding to the high-temperature regime, $V_B,\rho_m^B,\mathcal{H}_B$, is fitted with a bilinear form \cite{tanaka_simple_2000} $X_B(T,P) = x_0 + x_1 P + (y_0 + y_1 P) T$ having four fitting parameters. The anomalous parameters $\Delta E$ and $\Delta V$ are the same fitted by Eq.~\ref{eq:two-states_S_highT}. 
By direct derivation of Eq.~\ref{eq:two-states_S_highT} and Eq.~\ref{eq:two-states_volume}-\ref{eq:two-states_enthalpy}, the isobaric thermal expansion coefficient $\alpha_P$, the isobaric specific heat $c_P$ and the isothermal compressibility $k_T$ are \cite{tanaka_simple_2000}:
\begin{align}
    \alpha_P &= \frac{1}{V}\left( \frac{\partial V}{\partial T} \right)_P = \alpha_P^B + \frac{\Delta V (\Delta E + P \Delta V)}{V k_B T^2} \langle \xi \rangle , \label{eq:two-states_aP}\\
    c_P &  = \frac{1}{N} \left(\frac{\partial \mathcal{H} }{ \partial{T} }\right)_P = c_P^B + \frac{(\Delta E + P \Delta V)^2}{k_B T^2} \langle \xi \rangle , \label{eq:two-states_cP}\\
    k_T &= -\frac{1}{V} \left(\frac{\partial V }{ \partial P} \right)_T = k_T^B + \frac{(\Delta V)^2}{V k_B T} \langle \xi \rangle \label{eq:two-states_kT} ,
\end{align}
where $N$ is the number of atoms.
For the background part of the response functions, $\alpha_P^B,c_P^B,k_T^B$, we fit the high-temperature region of each isobar with a quadratic form \cite{tanaka_simple_2000} $X_B(T)|_P = a(P)+b(P)T+c(P)T^2$ except for $P=-1.5$ GPa. At the latter pressure we employ a linear form $X_B(T)|_P = a(P)+b(P)T$, since there are only three high-temperature data points.

The results of the TS fit are described in the main text, in Fig.\ref{fig:anomaly isobars} and in Table S5.

\subsection*{Coexistence line calculation}
We compute the coexistence line for the liquid and A7 crystal, and for the liquid and A17 crystal shown in Fig. 1 using the AIMLP of Ref. \cite{dragoni_mechanism_2021}. We first find a single coexistence temperature at a given pressure via the direct coexistence method in the NPT ensemble and then we integrate the coexistence line by the Clausius-Clapeyron equation \cite{kofke1993direct}. To estimate the direct coexistence at a given pressure we first relax the bulk liquid and crystal at different temperatures and fixed pressure in NPT. We preserve the symmetry of the crystals using an orthorhombic box and an hexagonal box for the A17 and A7 phase, respectively. We then build the coexistence box by superimposing the liquid and solid box, preserving the mean density of both liquid and crystal. We relax the coexistence box at fixed crystal particles, applying a barostat only along the orthogonal direction to the surface. Finally we run the coexistence simulations without the constraint on the crystal particles, repeating the simulation for $10$ independent replicas at each temperature. We estimate the melting temperature by studying the growth of the interface: below the melting temperature $T_m$, more than the $50\%$ of the $10$ runs crystallize, while above $T_m$ the majority of the runs melt.

\subsection*{Thermodynamic integration}
An essential step to compute the configurational entropy is to determine the absolute value of the free energy, thus the entropy, of the liquid at one state point.
We compute the liquid free energy at 900 K using equilibrium Thermodynamic Integration (TI) with respect to a Lennard-Jones (LJ) fluid. The theoretical details are specified in Appendix~E of the SI. We choose the LJ state point $T^*=1.3$, $\rho^*=0.70$ (in internal LJ units), since its radial distribution $g(r)$ is similar to the one of our system; the corresponding values for the LJ interaction are $\sigma=2.915$ $\Angstrom$ and $\varepsilon=59.66$ meV or $692.4$ K, and the thermal de Broglie wavelength is $0.05274~\Angstrom$. We cut and shift the LJ potential energy at $r_\mathrm{cut}^*=4$ in internal LJ units.

We employ the same method to measure the free energy of the A7 and A17 crystals at the same density at 300 K. In this case, the reference system is an Einstein crystal with spring constant 5 eV$/\Angstrom$.

\subsection*{Visualization softwares}
The drawings of crystal structures in Fig.\ref{fig:crystallization and crystals}c  are realized with the open source VESTA \cite{VESTA} software. The snapshots of crystals nucleated from the liquid in Fig.\ref{fig:crystallization and crystals}b are produced with the free version of the Ovito \cite{ovito} software.

\begin{acknowledgments}
The authors gratefully acknowledge funding from the PRIN 2020 project “Neuromorphic devices based on chalcogenide heterostructures” funded by the Italian Ministry for University and Research (MUR). This work has also been supported by the ICSC—Centro Nazionale di Ricerca in High Performance Computing, Big Data, and Quantum Computing funded by European Union—NextGenerationEU. The authors acknowledge the CINECA award under the ISCRA and ICSC initiatives for the availability of high performance computing resources and support.
\end{acknowledgments}

\bibliography{bibliography}

\end{document}